\documentclass[10pt,twocolumn,letterpaper]{article}

\usepackage{cvpr}              
\usepackage{amsmath}
\usepackage{xcolor}  
\usepackage{colortbl}
\usepackage{multirow}
\usepackage{latexsym}
\usepackage{tikz}
\usepackage{pifont}
\newcommand*\circled[1]{\tikz[baseline=(char.base)]{
            \node[shape=circle,draw,inner sep=0.7pt] (char) {#1};}}
\definecolor{cvprblue}{rgb}{0.21,0.49,0.74}
\usepackage[pagebackref,breaklinks,colorlinks,allcolors=cvprblue]{hyperref}

\def\paperID{} 
\def\confName{CVPR}
\def\confYear{2026}

\title{StructDiff: Structure-aware Diffusion Model for 3D Fine-grained \\ Medical Image Synthesis}

\author{Jiahao Xia$^{1,}$\footnotemark[1], \quad Yutao Hu$^{1,}$\footnotemark[1], \quad Yaolei Qi$^{1}$, \quad Zhenliang Li$^{1}$, \quad Wenqi Shao$^{4}$, \quad Junjun He$^{4}$, \\ Ying Fu$^{5}$, \quad Longjiang Zhang$^{6}$\footnotemark[2], \quad Guanyu Yang$^{1,2,3,}$\footnotemark[2]\\
$^{1}$Key Laboratory of New Generation Artificial Intelligence Technology and Its Interdisciplinary \\ Applications (Southeast University), Ministry of Education, China\\
$^{2}$Jiangsu Province Joint International Research Laboratory of Medical Information Processing, \\ Southeast University, Nanjing, China\\
$^{3}$Univ Rennes, CHU Rennes, Inserm, LTSI– UMR 1099, F-35000 Rennes, France \\
$^{4}$Shanghai AI Laboratory\\
$^{5}$School of Computer Science and Technology, Beijing Institute of Technology, Beijing, China\\
$^{6}$Department of Radiology, Jinling Hospital, Affiliated Hospital of Medical School, \\ Nanjing University, Nanjing, China}

\begin{document}
\maketitle

\footnotetext[1]{Equal contribution.}
\footnotetext[2]{Corresponding author.}

\begin{abstract}
Solving medical imaging data scarcity through semantic image generation has attracted growing attention in recent years. However, existing generative models mainly focus on synthesizing whole-organ or large-tissue structures, showing limited capability in reproducing fine-grained anatomical details. Due to the stringent requirement of topological consistency and the complex 3D morphological heterogeneity of medical data, accurately reconstructing fine-grained anatomical details remains a significant challenge. To address these limitations, we propose StructDiff, a Structure-aware Diffusion Model for fine-grained 3D medical image synthesis, which enables precise generation of topologically complex anatomies. In addition to the conventional mask-based guidance, StructDiff further introduces a paired image–mask template to guide the generation process, providing structural constrains and offering explicit knowledge of mask-to-image correspondence. Moreover, a Mask Generation Module (MGM) is designed to enrich mask diversity and alleviate the scarcity of high-quality reference masks. Furthermore, we propose a Confidence-aware Adaptive Learning (CAL) strategy based on Skip-Sampling Variance (SSV), which mitigates uncertainty introduced by imperfect synthetic data when transferring to downstream tasks. Extensive experiments demonstrate that StructDiff achieves state-of-the-art performance in terms of topological consistency and visual realism, and significantly boosts downstream segmentation performance. Code will be released upon acceptance.


\end{abstract}
\vspace{-2mm}
    
\section{Introduction}
\label{sec:intro}
\begin{figure}[t]
  \centering
   \includegraphics[width=\linewidth]{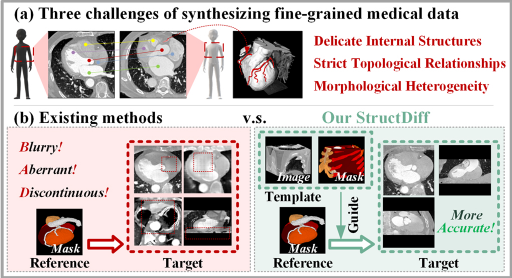}
   \caption{Synthesizing fine-grained 3D anatomical structures remains challenging due to three key factors. Consequently, existing methods often produce blurry and anatomically inconsistent results, whereas StructDiff leverages paired image–mask templates to generate more accurate and topologically coherent volumes.}
   \label{fig1}
   \vspace{-6mm}
\end{figure}

In recent years, the rapid advancement of artificial intelligence (AI) has created unprecedented opportunities to improve the efficiency and accuracy of medical image analysis. However, the success of AI in smart healthcare heavily depends on large-scale, high-quality annotated datasets~\cite{hu2023beyond}. Obtaining such voxel-level annotations is extremely expensive and time-consuming, especially for segmentation tasks~\cite{dayarathna2024deep, zhao2019data, hu2024omnimedvqa}. This challenge becomes even more pronounced in fine-grained anatomical structures, where the topology is complex and exhibits significant inter-individual variability~\cite{sengupta2024challenges, he2023geometric, qi2021examinee}. For instance, precise delineation of cardiac data requires expert knowledge and is difficult to scale, motivating the need for advanced generative models capable of synthesizing high-quality fine-grained medical data to augment training and facilitate downstream learning tasks.

Although recent diffusion-based generative models have achieved remarkable success in natural image synthesis, their direct application to medical imaging remains non-trivial due to the high dimensionality, structural complexity, and topological constraints inherent to medical data~\cite{dayarathna2024deep, nie2018medical}. Existing methods have made progress in synthesizing realistic images for 2D medical slices or coarse 3D organ-level structures, yet they fail to capture the intricate details of small-scale anatomical components. When applied to fine-grained regions that require high spatial precision, they often produce unrealistic or anatomically inconsistent results.

To better illustrate the challenges, we take cardiac computed tomography angiography (CTA) as a representative example. As shown in \cref{fig1}, the synthesis of cardiac images remains extremely difficult due to three key factors: (1) \textbf{Delicate internal structures}: The heart contains numerous thin coronary arteries and sub-vascular branches whose morphology is easily distorted during generation, leading to the loss of critical anatomical information. (2) \textbf{Strict topological relationships}: Cardiac anatomy exhibits highly constrained spatial organization among chambers and vessels~\cite{shi2022xmorpher}. Unguided generation typically results in structural discontinuities or unrealistic connections. (3) \textbf{Morphological heterogeneity}: Inter-patient and pathological variations cause large shape and texture differences across individuals~\cite{he2025homeomorphism}, further complicating fine-grained synthesis.

To overcome the aforementioned challenges and achieve high-quality synthesis of cardiac CTA images, we propose a Structure-aware Diffusion Model (StructDiff) for 3D fine-grained medical image generation. Besides the mask as a generation reference, StructDiff introduces a paired image–mask template as additional structural guidance. By doing so, the model establishes the explicit mask-to-image correspondence and preserves fine-grained topology, thereby improving the anatomical fidelity of the synthesized image. Meanwhile, the difficulty of annotating fine-grained volumetric data and the scarcity of annotated samples greatly limit the scalability of medical data generation. To mitigate this problem, a Mask Generation Module (MGM) is designed to increase mask diversity and supply topology-preserved reference masks, thereby enabling scalable synthesis in data-scarce scenarios. 

Moreover, due to the intricate structural composition of fine-grained anatomical regions, generating a flawless medical volume remains extremely challenging. To mitigate the impact of such imperfections and enhance the usability of synthetic data for downstream applications, we propose a Confidence-aware Adaptive Learning (CAL) strategy. During the diffusion process, Skip-Sampling Variance (SSV) is computed to estimate the uncertainty of each generated voxel, forming a voxel-level confidence map that quantifies synthesis reliability. When transferring to downstream pre-training, the confidence map serves as a dynamic weight to adaptively adjust the voxel-wise contribution in the loss computation, allowing the model to emphasize more reliable regions and suppress noisy ones. Overall, extensive experiments demonstrate that StructDiff substantially improves the quality of synthesized cardiac images and yields significant performance gains in downstream segmentation pre-training, confirming the effectiveness of StructDiff in 3D fine-grained medical image generation.

The key contributions of this work are:

\begin{itemize}
\setlength{\itemsep}{0.4em}
    \item We propose StructDiff for 3D medical image synthesis. To the best of our knowledge, this is the first approach that enables the generation of fine-grained, high-quality medical images while preserving topological consistency, which effectively alleviates the scarcity of medical data.

    \item We introduce a paired image–mask template as additional guidance to provide explicit mask-to-image correspondence information during generation and enhance anatomical fidelity. Meanwhile, MGM is designed to enrich the structural priors available to the generator.

    \item We develop Confidence-aware Adaptive Learning (CAL) strategy to mitigate the influence of imperfect synthesized medical data when transferring to downstream tasks. The CAL strategy incorporates Skip-Sampling Variance (SSV) to estimate voxel-level uncertainty and generate confidence maps, which adaptively adjust the voxel-wise contribution in the loss computation.
    
    \item Extensive experiments show that StructDiff achieves superior synthesis quality and significant improvements in downstream segmentation performance.
\end{itemize}

\section{Related Work}
\label{sec:related}
\subsection{Semantic Image Synthesis}
\label{subsec:sis}
Semantic image synthesis utilizes mask semantics and spatial information to guide the generative model, producing diverse images that inherently include segmentation masks, thereby offering valuable annotated data for downstream segmentation tasks.  The pioneering pix2pix\cite{isola2017image} framework establishes a foundational paradigm for mask-to-image mapping using Generative Adversarial Networks (GANs). Subsequent advancements\cite{park2019semantic} further refine semantic representation capacity through enhanced network architectures. Tan \textit{et al.}\cite{tan2021diverse} and Wang \textit{et al.}\cite{wang2021image} later achieve breakthroughs by analyzing semantic probability distributions and decoupling semantic content from style information, showing considerable improvements across multiple tasks. More recent explorations\cite{jeong20223d,lv2024place,zeng2023scenecomposer,tan2021efficient,berrada2024unlocking,liu2024swinit} have expanded research frontiers into 3D scene reconstruction, object arrangement optimization, model efficiency enhancement, and Transformer-based architectures.

The advent of diffusion models\cite{ho2020denoising}, characterized by their iterative denoising generation process, has showcased formidable generative capabilities that bolster semantic image synthesis tasks. Some works such as ControlNet\cite{zhang2023adding}, InstantID\cite{wang2024instantid}, T2I-Adapter\cite{mou2024t2i} and StableSketching\cite{koley2024s} demonstrate that incorporating rich prompts, such as Canny edges, segmentation masks, and sketches enables generation of high-quality images that meet specific requirements. 

However, directly applying these methods to medical images remains difficult, often producing anatomically inconsistent or distorted details when generating fine-grained anatomical structures, such as cardiac regions.

\subsection{Medical Image Generation}
\label{subsec:mig}
Medical image generation coupled with synthetic data-based pre-training represents a crucial paradigm for addressing data scarcity in medical imaging. Recent advancements have explored various generative approaches including unconditional synthesis, conditional generation, and multi-modal fusion, aiming to produce diverse and realistic medical images. Due to heterogeneous acquisition difficulties across imaging modalities, researchers have developed cross-modal conversion techniques\cite{peng2024cbct, lyu2023generative} and super-resolution reconstruction algorithms \cite{kim2024adaptive, huang2017simultaneous}. Notably, several studies\cite{jung2021conditional,aversa2023diffinfinite,osuala2024pre,chen2024towards} claim that incorporating attention maps, class labels, and tumor masks as conditional inputs significantly improves generation quality compared to unguided methods, particularly in localized synthesis.

Müller-Franzes \textit{et al.}\cite{muller2023multimodal} experimentally validate that slower synthesis speed of  diffusion models is justified by superior image quality in medical contexts. Han \textit{et al.}\cite{han2023medgen3d} achieve 3D reconstruction by aggregating multi-view 2D slices. Meanwhile, other studies\cite{xu2024medsyn,wang20243d,dorjsembe2024conditional,guo2024maisi} pioneer high-resolution 3D training through various resizing or cropping strategies. Moreover, some researchers\cite{kreitner2024synthetic,khader2023denoising,kamran2021vtgan,wang2024self} have further advanced this field by demonstrating that synthetic medical images can promote the performance of downstream tasks particularly when real data is scarce.

However, existing methods still struggle to synthesize fine-grained 3D medical images with complex anatomical structures and strict topological constraints. To address this issue, we propose StructDiff that incorporates paired image–mask templates to establish the explicit mask-to-image correspondence and provides fine-grained topological information. This design enables anatomically faithful and high-quality medical image generation beyond prior approaches.
\section{Methodology}
\label{sec:method}
\begin{figure*}[t]
  \centering
   \includegraphics[width=0.95\textwidth]{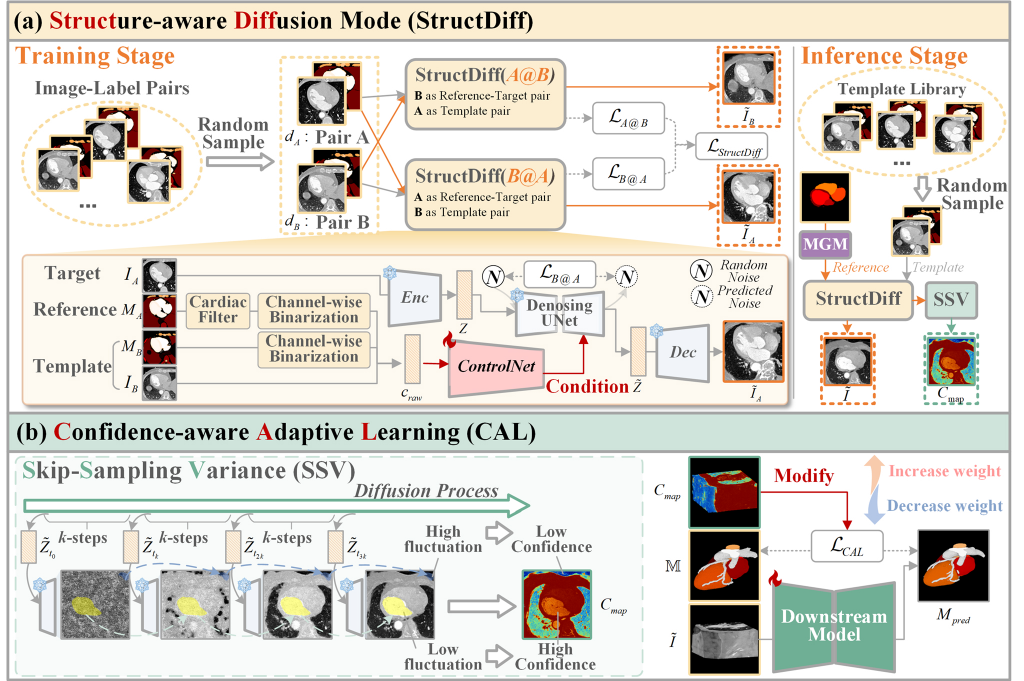}

   \caption{Overview of the proposed framework. (a) The Structure-aware Diffusion Model (StructDiff) is designed to generate the precise, diverse, and topology-preserved medical images based on template-guided conditions. (b) The Confidence-aware Adaptive Learning (CAL) strategy facilitates downstream segmentation pre-training by reducing the effect of imperfect synthetic samples.}
   \label{fig2}
   \vspace{-4mm}
\end{figure*}

This section presents a comprehensive overview of the proposed framework. As shown in \cref{fig2}, our work comprises two principal components, StructDiff and CAL strategy. We will elaborate on these two important components in detail in the following parts.

\subsection{Structure-aware Diffusion Model}
\label{3.1}
\subsubsection{Semantic Reference with Template Guidance}
\label{3.1.1}
During the training phase, two data pairs $\boldsymbol{d}_{A}=\{\boldsymbol{I}_{A},\boldsymbol{M}_{A}\}$ and $\boldsymbol{d}_{B}=\{\boldsymbol{I}_{B},\boldsymbol{M}_{B}\}$ are sampled from the training set $\boldsymbol{D}^{tr}=\{\boldsymbol{d}_{l}^{tr}\}_{l=1}^{{N}^{tr}}$. After spatial resampling, all volumes $\boldsymbol{I}_{A}$, $\boldsymbol{M}_{A}$, $\boldsymbol{I}_{B}$, $\boldsymbol{M}_{B}\in \mathbb{R}^{D\times H\times W}$ are standardized to the same size and then fed into the generator as inputs, adopting two distinct patterns: $A @ B$ and $B @ A$, respectively. Taking the generation direction $B @ A$ as an example. $\boldsymbol{M}_B$ and $\boldsymbol{I}_B$ are employed as the template pair to provide the mask-to-image correspondence. $\boldsymbol{M}_A$ is utilized as the reference while $\boldsymbol{I}_A$ is regarded as the learning target. Specifically, $\boldsymbol{M}_{A}$ is filtered to exclude coarse-grained regions, thereby enabling the model to prioritize fine-grained structures (\emph{e.g.} coronary arteries, ascending aorta, left atrium,  right ventricle, \emph{etc}). In this way, references provide fine-grained topological information to ensure structural stability of the target.

Ensuring adequate mask representation while avoiding computational overcost, inspired by MAISI \cite{guo2024maisi}, we implement channel-wise binarization for masks $\boldsymbol{M}_{A}$ and $\boldsymbol{M}_{B}$ to obtain $\boldsymbol{M}_{A}^{'}$ and $\boldsymbol{M}_{B}^{'}$. We then concatenate them with the template image $\boldsymbol{I}_{B}$ to form the raw condition feature $\boldsymbol{\mathrm{c}}_{raw}=\boldsymbol{M}_{A}^{'}\circled{C}\boldsymbol{M}_{B}^{'}\circled{C}\boldsymbol{I}_{B} \in \mathbb{R}^{C\times D\times H\times W}$, where $\circled{C}$ denotes the concatenation along the channel axis. Then $\boldsymbol{\mathrm{c}}_{raw}$ is encoded by a trainable ControlNet $\boldsymbol{E}_{ctrl}$ and then utilized to guide the synthesis process of $\boldsymbol{I}_{A}$. The process can be defined as $\boldsymbol{\mathrm{c}}=\boldsymbol{E}_{ctrl}\left(\boldsymbol{\mathrm{c}}_{raw}\right)$.

For the generation target $\boldsymbol{I}_{A}$, we first obtain the latent space encoding feature $\boldsymbol{Z}\in \mathbb{R}^{4\times\frac{D}{4}\times\frac{H}{4}\times\frac{W}{4}}$ via a pre-trained encoder with frozen parameters. The diffusion model iteratively adds random noise by $\boldsymbol{Z}_{t}=\sqrt{1-\beta_{t}}\boldsymbol{Z}_{t-1}+\sqrt{\beta_{t}}\boldsymbol{\epsilon}_{t}$ to obtain $\boldsymbol{Z}_{T}\sim\mathcal{N}(0,I)$, where $\beta_t$ is used to control the diffusion process. 
Then we sample $\widetilde{\boldsymbol{Z}}_{T}$ from pure Gaussian noise and sample $\widetilde{\boldsymbol{Z}}_{0}$ by denoising following \cref{eq6}. 
\vspace{-2mm}
\begin{equation} 
\resizebox{0.9\linewidth}{!}{
$
{p}_{\boldsymbol{\theta}}(\boldsymbol{\widetilde{Z}}_{t-1}|\boldsymbol{\widetilde{Z}}_{t})=\mathcal{N}(\boldsymbol{\widetilde{Z}}_{t-1};\boldsymbol{\mu_{\boldsymbol{\theta }}}(\boldsymbol{\widetilde{Z}}_{t},\boldsymbol{t},\boldsymbol{c}),\boldsymbol{\Sigma_{\boldsymbol{\theta }}}(\boldsymbol{\widetilde{Z}}_{t},\boldsymbol{t},\boldsymbol{c})).
$
}
\label{eq6}
\vspace{-2mm}
\end{equation}

Using condition feature $\boldsymbol{\mathrm{c}}$ as guidance, we integrate time embedding $\boldsymbol{t}$ and current state $\boldsymbol{\widetilde{Z}}_{t}$ into a denoising model to predict noise $\boldsymbol{\widetilde{\epsilon}}_{t}$. Subsequently, we compute the distance between $\boldsymbol{\widetilde{\epsilon}}_{t}$ and real noise $\boldsymbol{\epsilon}_{t}$ as the training objective:
\vspace{-2mm}
\begin{equation} 
\mathcal{L}_{DM}=\mathbb{E}_{\widetilde{\boldsymbol{Z}}_{0},\boldsymbol{\epsilon}_{t}\sim\mathcal{N}\left(0,1\right),\boldsymbol{t},\boldsymbol{\mathrm{c}}}\left[{\left\Vert\boldsymbol{\epsilon}_{t}-\boldsymbol{\widetilde{\epsilon}}_{t}\left(\boldsymbol{\widetilde{Z}}_{t},\boldsymbol{t},\boldsymbol{\mathrm{c}}\right)\right\Vert}^{2}_{2}\right],
\label{eq2}
\vspace{-2mm}
\end{equation}
where $\widetilde{\boldsymbol{Z}}_{0}$ represent the clean latent feature at step $0$ of the reverse diffusion process. Then $\widetilde{\boldsymbol{Z}}_{0}$ will be decoded into synthetic Image $\boldsymbol{\widetilde{I}}_A$ by a decoder. 

\subsubsection{Bidirectional Training Strategy}
\label{3.1.2}
Our proposed template-guided model necessitates the simultaneous loading of two data pairs $\boldsymbol{d}_{A}=\{\boldsymbol{I}_{A},\boldsymbol{M}_{A}\}$ and $\boldsymbol{d}_{B}=\{\boldsymbol{I}_{B},\boldsymbol{M}_{B}\}$. Therefore, to enrich the training combination, we introduce a bidirectional training strategy, where $\boldsymbol{d}_{A}$ and $\boldsymbol{d}_{B}$ alternately serve as a template for the other by turns. Specifically, as shown in \cref{fig2}(a), $A @ B$ and $B @ A$ represent that using $\boldsymbol{d}_{A}$ as the template pair to generate $\boldsymbol{I}_{B}$ and using $\boldsymbol{d}_{B}$ as the template pair to generate $\boldsymbol{I}_{A}$, respectively. Using bidirectional training strategy, we derive the final loss function of StructDiff as expressed:
\vspace{-2mm}
\begin{equation} 
\mathcal{L}_{StructDiff}=\mathcal{L}_{{DM}_{A @ B}} + \mathcal{L}_{{DM}_{B @ A}}.
\label{eq5}
\vspace{-2mm}
\end{equation}

By integrating template-guided conditioning through bidirectional training strategies, we generate $(N^{tr})^2$ training sample pairs solely based on $N^{tr}$ image-mask pairs through random sampling and combinations, which significantly reduces the amount of data required for training the generator compared to previous methods\cite{guo2024maisi,dorjsembe2024conditional}.

\begin{figure}[t]
  \centering
   \includegraphics[width=\linewidth]{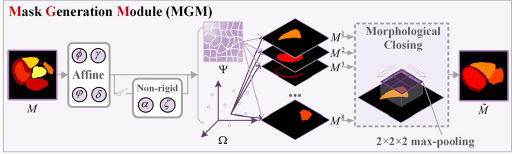}
   \caption{The workflow of Mask Generation Module.}
   \label{fig3}
   \vspace{-5mm}
\end{figure}
\subsubsection{Training-free Mask Generation Module}
\label{3.1.3}
Having more diverse masks indicates more morphologically heterogeneous images to be generated. Consequently, as shown in \cref{fig3}, we design MGM to produce additional masks with stringent structure during the inference stage.

Specifically, in the inference stage, we select a small subset $\boldsymbol{Temp}=\{\boldsymbol{d}^{T}_{l}\}^{N^T}_{l=1}$ from the test dataset $\boldsymbol{D}^{ts}=\{\boldsymbol{d}_{l}^{ts}\}_{l=1}^{{N}^{ts}}$ to create a template library that provides random template pairs for subsequent inference. We randomly choose a mask $\boldsymbol{M}$ from $\boldsymbol{Temp}$ and exclude its coarse-grained region. Then, $\boldsymbol{M}$ undergo an affine and an non-rigid deformation to generate the registration filed matrix $\Omega$ and the deformed coordinates $\Psi$ as shown in \cref{eq3}:
\vspace{-2mm}
\begin{equation} 
\boldsymbol{\Omega},\boldsymbol{\Psi} = \mathcal{T}_{nr}\left(\mathcal{T}_{a}\left(\boldsymbol{M},\phi,\gamma,\varphi,\delta\right),\alpha,\zeta\right),
\label{eq3}
\vspace{-2mm}
\end{equation}
where $\phi$,$\gamma$,$\varphi$, and $\delta$ control the affine transformation $\mathcal{T}_{a}$, as well as $\alpha$ and $\zeta$ regulate the non-rigid deformation $\mathcal{T}_{nr}$.

The discretization of voxel coordinates caused by deformation results in spatially disconnected micro-fractures in the deformed masks, thereby compromising the inherent spatial coherence of masks. This discontinuity adversely affects subsequent image generation tasks, particularly in preserving the integrity of coronary arteries. To address this challenge, we utilize a max-pooling operator with a kernel size of $2\times2\times2$ to perform morphological closing on each class of deformed mask, achieving the continuous mask $\boldsymbol{\widetilde{M}}$:
\vspace{-2mm}
\begin{equation} 
\resizebox{!}{0.35cm}{$
\boldsymbol{\widetilde{M}} = \bigcup^{8}_{i=1}\mathrm{Maxpool}_{2\times2\times2}(\Lambda(\boldsymbol{M^i},\boldsymbol{\Omega},\boldsymbol{\Psi})), $}
\label{eq4}
\vspace{-2mm}
\end{equation}
where $\Lambda$ represents the deformation function, and $\boldsymbol{M}^{i}$ denotes the portion of $\boldsymbol{M}$ that includes only the ${i}^{th}$ class mask. 

It is worthwhile to mention that our MGM operates using a training-free strategy, which does not require training an additional generative network and functions in a computationally efficient manner. By doing so, the MGM provides StructDiff with a more diverse set of masks that satisfy both topological consistency and preservation of fine-grained details, enabling the generation of large-scale synthetic datasets for downstream pre-training tasks.

\vspace{-1mm}
\subsection{Confidence-aware Adaptive Learning}
\vspace{-1mm}
\label{3.2}
According to \cref{eq6}, diffusion models are based on Hidden Markov Chains and sampling at each step is contingent upon the current state\cite{ho2020denoising}. As shown in \cref{fig2}, we incorporate the SSV estimation during the reverse sampling process to obtain a Confidence Map, which quantifies the reliability of the imperfect synthetic images and are utilized to guide the down-stream pre-training.

\subsubsection{Skip-Sampling Variance Estimation}
\label{3.2.1}
Throughout the T-step reverse diffusion, once the model has sufficiently learned the distribution of a region, the sampling process for that region tends to be stable, leading to smaller variations across steps. In contrast, for regions where the model has not yet achieved robust learning, the reverse sampling process exhibits greater uncertainty and fluctuations in these areas.

According to the above analysis, we design the SSV with regard to the efficiency of the diffusion model. Specifically, it takes $k$-step as intervals to perform staggered sampling over the $T$-step denoising process, resulting in $T'=\{t_0,t_k,\dots,t_{rk}\}$, where ${r\cdot k}\leq {T-1}$. For each $t\in T'$, we obtain the latent space feature $\boldsymbol{\widetilde{Z}}_t$ and decode it to generate $\boldsymbol{\widetilde{I}}_t$~\cite{stracke2025cleandift}. Consequently, we can derive a set of intermediate images represented by $\boldsymbol{\Pi=\{\boldsymbol{\widetilde{I}}_{t_{0}},\boldsymbol{\widetilde{I}}_{t_{k}},\dots,\boldsymbol{\widetilde{I}}_{t_{rk}}\}}$. Then we generate a confidence map $\boldsymbol{C}_{map}$ as follows:
\vspace{-2mm}
\begin{equation} 
\resizebox{!}{0.44cm}{$
\boldsymbol{C}_{map}=1-\mathrm{Norm}(\frac{\Sigma^r_{i=0}(\boldsymbol{\widetilde{I}}_{t_{ik}}-\overline{\boldsymbol{\Pi}})^{2}}{r}),
$}
\label{eq7}
\vspace{-2mm}
\end{equation}
where $\overline{\boldsymbol{\Pi}}$ denotes the average value of the image set $\boldsymbol{\Pi}$, and the function $\mathrm{Norm}(\cdot)$ represents the normalization of the content. As $\boldsymbol{C}_{map}$ reflects the stability of the generation process and the certainty of synthetic images, higher values indicate stable generation, implying higher-quality synthetic data. In contrast, lower values of $\boldsymbol{C}_{map}$ correspond to the imperfect result with lower reliability, which should be avoid when transferring to pre-training.

\subsubsection{Using Confidence Map to Enhance Pre-training}
\label{3.2.2}
For mask $\mathbb{M}$ (either directly providing $\boldsymbol{M}$ or deriving $\boldsymbol{\widetilde{M}}$ from MGM), we synthesize its corresponding image $\boldsymbol{\widetilde{I}}$ and confidence map $\boldsymbol{C}_{map}$ through StructDiff and SSV, thereby forming a synthetic dataset entry $\boldsymbol{d}_{syn}=\{\mathbb{M},\boldsymbol{\widetilde{I}},\boldsymbol{C}_{map}\}$.
Most prior ``synthesis then pre-training'' approaches, as shown in \cref{eq8}, directly calculate the gradient for the each voxel without differentiation, where $F$ refers to the downstream model and $\mathcal{L}_{ori}$ denotes the original loss function. 
\vspace{-2mm}
\begin{equation}
\resizebox{!}{0.4cm}{$
\mathcal{L}_{{ori}}=\mathcal{L}\left(F\left(\boldsymbol{\widetilde{I}}\right),\mathbb{M}\right). $}
\label{eq8}
\vspace{-2mm}
\end{equation}

In contrast, CAL leverages the generated confidence map to promote the downstream segmentation pre-training according to \cref{eq9}, which dynamically adjusts the contribution of each voxel during the loss computation: 
\vspace{-2mm}
\begin{equation}
    \mathcal{L}_{CAL}=\boldsymbol{C}_{map} \cdot \mathcal{L}\left(F\left(\boldsymbol{\widetilde{I}}\right), \mathbb{M}\right).
\label{eq9}
\vspace{-2mm}
\end{equation}

In this way, CAL assigns greater weight to confident regions and mitigates the influence of uncertain regions in imperfectly synthesized images, thereby yielding more accurate pre-training.

\section{Experiments}
\label{sec:experiments}

\begin{table*}[t]
    \centering
    \caption{Comparison with state-of-the-art methods in terms of generative evaluation metrics. \colorbox{gray!20}{\textbf{Bold}} denotes the best results, and \underline{Underlined} denotes the best results among 2D generative methods. ``-'' denotes the metrics is unavailable for this method.}
    \footnotesize
    \begin{tabular}{@{}c|l|c|c|c|c|c|c}
        \toprule
         \multicolumn{2}{c|}{\textbf{Methods}} & \textbf{Dim} & \textbf{Type} & \textbf{SSIM$\uparrow$} & \textbf{RMSE$\downarrow$} & \textbf{LPIPS$\downarrow$} & \textbf{FID$\downarrow$}  \\
        \midrule
         \multirow{3}{*}{Unconditional} &HA-GAN\cite{sun2022hierarchical} & 3D& GAN& - & -  & -& 6.3975\\
         ~&WDM\cite{friedrich2024wdm} & 3D& DM& - & - & - & 5.2148\\
         ~&Medical Diffusion\cite{khader2023denoising} & 3D& DM& - & - & - & 7.1284\\
        \midrule
         \multirow{7}{*}{With Surroundings} &SPADE\cite{park2019semantic} & 2D& GAN& 0.8442$\pm$0.0152 & 0.1812$\pm$0.0364 & 0.0855$\pm$0.0512 & 2.9331\\
         &ControlNet\cite{zhang2023adding} & 2D& DM&\underline{0.8739$\pm$0.0197}& \underline{0.1182$\pm$0.0436}& 0.0685$\pm$0.0324& 2.0073 \\
         &T2I-Adapter\cite{mou2024t2i}  & 2D&DM& 0.8641$\pm$0.0204 & 0.1272$\pm$0.0488 & \underline{0.0621$\pm$0.0300}& \underline{1.9812} \\
         &DiffInfinite\cite{aversa2023diffinfinite} & 2D &DM& 0.4076$\pm$0.0368 & 0.6984$\pm$0.3289  & 0.5943$\pm$0.3725& 7.4952\\
         &pix2pix-3D\cite{isola2017image}  & 3D&GAN& 0.7499$\pm$0.0263 & 0.3706$\pm$0.1043 & 0.1366$\pm$0.0821  & 3.9607\\
         &Med-DDPM\cite{dorjsembe2024conditional} & 3D &DM& 0.8561$\pm$0.0153 & 0.2437$\pm$0.0844  & 0.1038$\pm$0.0549& 3.2129\\
         &MAISI\cite{guo2024maisi}  & 3D&DM& 0.9110$\pm$0.0158 & 0.0836$\pm$0.0141 & 0.0553$\pm$0.0152 & 1.6881 \\ 
        \midrule
         \multirow{8}{*}{W/o Surroundings}&SPADE\cite{park2019semantic} & 2D &GAN& 0.8233$\pm$0.0161 & 0.2283$\pm$0.0764 & 0.1045$\pm$0.0528 & 3.7782  \\
         &ControlNet\cite{zhang2023adding} & 2D &DM& 0.8342$\pm$0.0231&0.1975$\pm$0.0683 & 0.0767$\pm$0.0282 & 2.3926 \\
         &T2I-Adapter\cite{mou2024t2i}  & 2D&DM& 0.8424$\pm$0.0281 & 0.2099$\pm$0.0692 &0.0724$\pm$0.0304 & 2.4233  \\
         &DiffInfinite\cite{aversa2023diffinfinite}  & 2D& DM&0.3874$\pm$0.0392 & 0.7098$\pm$0.3675 & 0.6039$\pm$0.4317 & 7.8165 \\
         &pix2pix-3D\cite{isola2017image} & 3D &GAN& 0.7272$\pm$0.0272 & 0.3921$\pm$0.1275& 0.1458$\pm$0.0983   & 4.1224 \\
         &Med-DDPM \cite{dorjsembe2024conditional} & 3D&DM& 0.8128$\pm$0.0168 & 0.3026$\pm$0.0817 & 0.1284$\pm$0.0893 & 3.7139\\
         &MAISI \cite{guo2024maisi} & 3D& DM&0.9072$\pm$0.0196 & 0.0954$\pm$0.0181 & 0.0572$\pm$0.0231 & 1.8214 \\
         \cline{2-8}
         &\textbf{StructDiff} (ours) & 3D & DM&\cellcolor{gray!20}\textbf{0.9243$\pm$0.0165} & \cellcolor{gray!20}\textbf{0.0731$\pm$0.0155} & \cellcolor{gray!20}\textbf{0.0541$\pm$0.0140} & \cellcolor{gray!20}\textbf{1.5821} \\
        \bottomrule
    \end{tabular}
    \label{tab1}
    \vspace{-3mm}
\end{table*}

\begin{figure*}[t]
  \centering
  \resizebox{0.98\textwidth}{!}{
   \includegraphics[width=\textwidth]{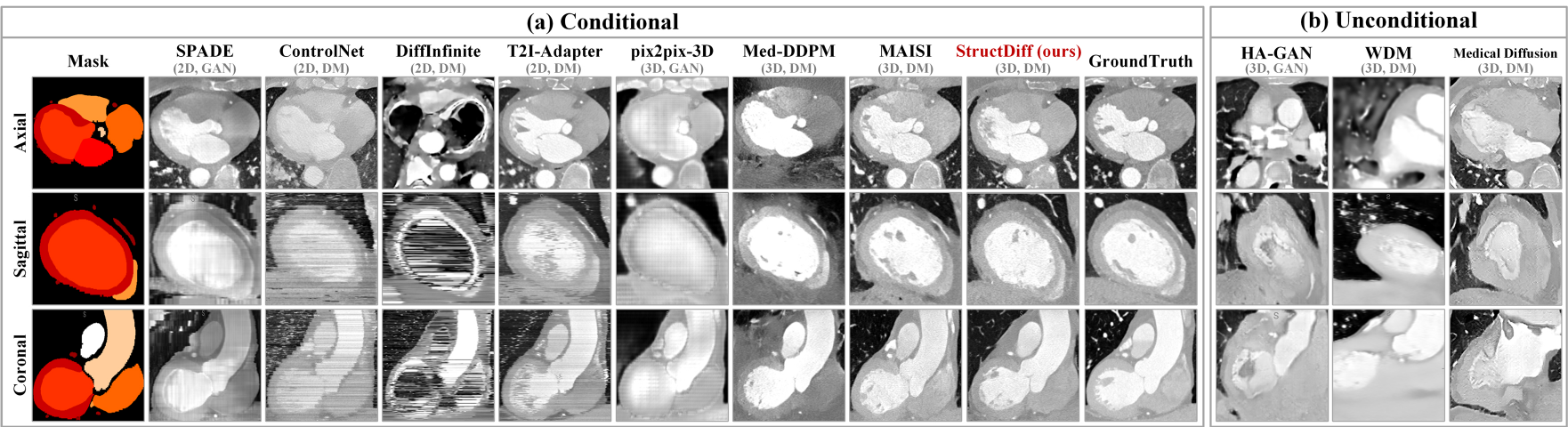}}
   \caption{Comparison of synthetic fine-grained cardiac images generated by existing methods and our StructDiff framework. (a) The results of the conditional generation methods based on the given mask.(b) The results of the unconditional generation methods.}
   \label{fig4}
\vspace{-5mm}
\end{figure*}

\begin{table*}[t]
\centering
\caption{Comparison of different methods in terms of pre-training 
 performance on downstream segmentation tasks. \colorbox{gray!20}{\textbf{Bold}} denotes the best results, and \underline{Underlined} denotes the best results among 2D generative methods.}
    \resizebox{0.90\textwidth}{!}{
    \begin{tabular}{@{}c|c|c|cc|cc|cc|cc@{}}
        \toprule
          \multicolumn{2}{c|}{\multirow{2}*{\textbf{Method}}} & \multirow{2}*{\textbf{Dim}}& \multicolumn{2}{c|}{\textbf{ImageALCAPA}\cite{zeng2022imagealcapa}} & \multicolumn{2}{c|}{\textbf{ImageCHD}\cite{xu2020imagechd}} & \multicolumn{2}{c|}{\textbf{MM-WHS}\cite{zhuang2019mmwhs} (7S)} & \multicolumn{2}{c}{\textbf{ASOCA}\cite{gharleghi2022automated}} \\
         \cline{4-5} \cline{6-7} \cline{8-9} \cline{10-11}
         \multicolumn{2}{c|}{~} & ~ & \textbf{Dice} $(\%)$$\uparrow$ & \textbf{RDice} $(\%)$ $\uparrow$& \textbf{Dice} $(\%)$$\uparrow$ & \textbf{RDice} $(\%)$$\uparrow$ & \textbf{Dice} $(\%)$$\uparrow$ & \textbf{RDice} $(\%)$$\uparrow$ & \textbf{Dice} $(\%)$$\uparrow$ & \textbf{RDice} $(\%)$$\uparrow$ \\
         
        \midrule
        
         \multirow{9}{*}{nnU-Net} &Real-Only & - &
         62.35$\pm$8.76 & 66.14$\pm$9.76 & 75.17$\pm$8.83& 79.34$\pm$9.94&
         78.63$\pm$2.42 & 80.88$\pm$2.58& 70.29$\pm$8.37& 74.58$\pm$9.42\\
         &SPADE \cite{park2019semantic} & 2D &
         60.12$\pm$10.37& 64.43$\pm$12.54& 76.85$\pm$9.31& 79.98$\pm$10.53&
         82.44$\pm$2.87 & 85.73$\pm$2.92& 64.67$\pm$10.15& 68.93$\pm$11.21\\
         &ControlNet \cite{zhang2023adding}& 2D &
         61.38$\pm$9.72& 66.84$\pm$9.61& \underline{78.09$\pm$8.42}& \underline{81.23$\pm$9.52}&
         85.17$\pm$2.93 & 88.39$\pm$3.01& \underline{78.21$\pm$8.19}& \underline{82.42$\pm$9.38}\\
         &T2I-Adapter \cite{mou2024t2i}& 2D &
         \underline{62.61$\pm$8.52}& \underline{67.02$\pm$9.43}& 77.58$\pm$8.63& 80.74$\pm$9.74&
         \underline{86.22$\pm$2.75} & \underline{89.40$\pm$2.87}& 77.83$\pm$8.03& 81.98$\pm$9.12\\
         &DiffInfinite \cite{aversa2023diffinfinite}& 2D &
         50.41$\pm$18.62& 54.53$\pm$20.74& 65.08$\pm$15.17& 69.32$\pm$17.24&
         72.05$\pm$6.84 & 75.23$\pm$6.90& 48.03$\pm$18.62& 52.22$\pm$20.78\\
         &pix2pix-3D \cite{isola2017image}& 3D &
         56.89$\pm$12.51& 61.02$\pm$13.66& 68.31$\pm$13.32& 72.52$\pm$14.43&
         72.61$\pm$7.46 & 75.79$\pm$7.54& 57.65$\pm$15.63& 61.82$\pm$16.73\\
         &Med-DDPM \cite{dorjsembe2024conditional}& 3D &
         62.87$\pm$8.29& 65.52$\pm$10.82& 74.29$\pm$8.03& 78.46$\pm$9.11&
         86.37$\pm$2.79 & 89.53$\pm$2.87& 75.53$\pm$11.72& 79.70$\pm$12.88\\
         &MAISI \cite{guo2024maisi}& 3D &
         63.42$\pm$8.11& 67.58$\pm$9.22& 78.79$\pm$7.76& 81.93$\pm$8.86&
         88.51$\pm$2.57 & 91.68$\pm$2.64& 81.22$\pm$8.63& 85.39$\pm$9.75\\
         &\textbf{StructDiff (ours)}&  3D &
         \cellcolor{gray!20}\textbf{64.09$\pm$8.31} & \cellcolor{gray!20}\textbf{68.26$\pm$9.43}& \cellcolor{gray!20}\textbf{79.05$\pm$7.83}& \cellcolor{gray!20}\textbf{82.21$\pm$8.97}&
         \cellcolor{gray!20}\textbf{89.31$\pm$2.76} & \cellcolor{gray!20}\textbf{92.48$\pm$2.83}& \cellcolor{gray!20}\textbf{82.23$\pm$7.85}& \cellcolor{gray!20}\textbf{86.41$\pm$8.91}\\
        \midrule
         \multirow{9}{*}{MedSAM}&Real-Only & - &
         58.49$\pm$3.72& 62.76$\pm$4.07& 66.21$\pm$5.83& 70.39$\pm$6.41&
         74.76$\pm$5.87 & 77.98$\pm$6.37& 67.32$\pm$11.42& 71.48$\pm$12.70\\
         &SPADE \cite{park2019semantic}& 2D &
         56.15$\pm$5.69& 60.32$\pm$6.42& 65.63$\pm$7.85& 69.78$\pm$8.72&
         73.28$\pm$6.29 & 76.45$\pm$6.71& 61.62$\pm$15.08& 65.79$\pm$16.33\\
         &ControlNet \cite{zhang2023adding}& 2D &
         59.58$\pm$2.75& 63.73$\pm$2.97& 71.07$\pm$5.06& 75.22$\pm$5.58&
         76.29$\pm$2.58 & 79.43$\pm$2.74& \underline{70.12$\pm$10.54}& \underline{74.27$\pm$11.51}\\
         &T2I-Adapter \cite{mou2024t2i}& 2D &
         \underline{60.07$\pm$2.76}& \underline{64.21$\pm$3.02}& \underline{71.33$\pm$5.35}& \underline{75.42$\pm$5.90}&
         \underline{76.61$\pm$2.39} & \underline{79.75$\pm$2.56}&69.77$\pm$10.03& 73.94$\pm$11.08 \\
         &DiffInfinite \cite{aversa2023diffinfinite}& 2D &
         50.12$\pm$9.43& 54.27$\pm$10.34& 56.14$\pm$15.72& 60.31$\pm$17.38&
         60.27$\pm$11.09 & 63.44$\pm$11.83& 50.83$\pm$17.85& 53.99$\pm$19.73\\
         &pix2pix-3D \cite{isola2017image}& 3D &
         53.31$\pm$7.76& 57.48$\pm$8.51& 59.72$\pm$13.18& 63.87$\pm$14.49&
         67.35$\pm$8.87 & 70.49$\pm$9.53& 58.41$\pm$17.06& 62.55$\pm$18.62\\
         &Med-DDPM \cite{dorjsembe2024conditional}& 3D &
         58.17$\pm$3.15& 62.31$\pm$3.48& 68.79$\pm$6.87& 72.93$\pm$7.65&
         74.93$\pm$3.41 & 79.08$\pm$3.57& 65.87$\pm$13.75& 70.01$\pm$14.92\\
         &MAISI \cite{guo2024maisi}& 3D & 59.71$\pm$2.68&63.86$\pm$2.95&71.42$\pm$5.24&75.55$\pm$5.82&
      77.38$\pm$2.13 &80.51$\pm$2.22&70.51$\pm$10.76&74.65$\pm$11.85\\
         &\textbf{StructDiff (ours)}& 3D &
         \cellcolor{gray!20}\textbf{60.63$\pm$2.65} & \cellcolor{gray!20}\textbf{64.80$\pm$2.89}& \cellcolor{gray!20}\textbf{71.92$\pm$4.97}&
         \cellcolor{gray!20}\textbf{76.05$\pm$5.51}&
         \cellcolor{gray!20}\textbf{78.75$\pm$1.29} & \cellcolor{gray!20}\textbf{81.89$\pm$1.33}& \cellcolor{gray!20}\textbf{71.83$\pm$9.73}&
         \cellcolor{gray!20}\textbf{76.01$\pm$10.71}\\
        \bottomrule
    \end{tabular}
    }
    \label{tab2}
    \vspace{-4mm}
\end{table*}

\subsection{Experiment Settings}
\label{4.1}
\subsubsection{Dataset}
\label{4.1.1}
\textbf{Generation stage}. We train generation models using ImageCAS~\cite{zeng2023imagecas}, which is a large-scale coronary segmentation dataset that comprises 1000 cases. Dataset is split into training and test sets at a ratio of 1:4. Among the test set, 50 cases are placed into the template library, while 750 cases are used as references. All images undergo initial segmentation using pretrained nnU-Net~\cite{isensee2021nnu} and TotalSegmentator~\cite{wasserthal2023totalsegmentator} to generate masks for seven cardiac substructures and surrounding anatomical tissues, respectively. 

\noindent\textbf{Pre-training stage}.
For downstream segmentation, we select four public and one private datasets for cardiac segmentation. Each dataset is split into training and test sets at a ratio of 1:4. The four public datasets are ImageCHD~\cite{xu2020imagechd}, ImageALCAPA~\cite{zeng2022imagealcapa}, ASOCA~\cite{gharleghi2022automated}, and MM-WHS~\cite{zhuang2019mmwhs}. The private dataset comprises 288 cases that include masks for coronary arteries and cardiac substructures. 

\subsubsection{Implementation Details}
\label{4.1.2}
Our experiments compare to a series of representative methods, including unconditional and conditional generation frameworks applied in 2D or 3D spaces. We utilize the MONAI framework\cite{gupta2024monai} for the implementation. In the generation stage, StructDiff is trained using the AdamW optimizer with an initial learning rate of $1e^{-5}$ for 200 epochs. Meanwhile, all data undergo resampling and are randomly cropped to a size of $512\times512\times128$, with a batch size of 4. During the inference phase, a sliding window is applied to recover latent features, outputting a 3D volume with a resolution of $512\times512\times256$. Additionally, StructDiff is trained on a single NVIDIA RTX A100 GPU, while inference is performed on a single NVIDIA RTX 4090 GPU.

In the pre-training stage, all downstream segmentation experiments are performed on a single NVIDIA RTX 4090 GPU, and all data are processed according to the default preprocessing procedures of nnU-Net\cite{isensee2021nnu} and MedSAM\cite{ma2024segment}. The ``Real-Only" experiments are trained on limited real data for 1000 epochs, while other methods train on synthetic datasets for 800 epochs first, followed by fine-tuning on limited real data for 200 epochs.

\subsubsection{Evaluation Metrics}
\label{4.1.3}
During the generation stage, we evaluate image quality using four metrics\cite{guo2024maisi}: Structural Similarity Index (SSIM), Root Mean Square Error (RMSE), Fréchet Inception Distance (FID), and Learned Perceptual Image Patch Similarity (LPIPS). In the pre-training stage, we use Dice and RDice \cite{qi2021examinee} as our metrics. The detailed explanation and calculation of metrics are presented in the supplementary materials.

\subsection{Generation Comparison}
\label{4.2}
\vspace{-1mm}
\subsubsection{Qualitative Evaluation}
\label{4.2.1}
As shown in \cref{fig4}, cardiac images generated by unconditional methods exhibit compromised topological consistency due to the lack of spatial guidance, whereas images produced by most conditional methods demonstrate preserved structural accuracy. Thus, sufficient guidance is significant for the synthesis of anatomy with fine-grained structure. Meanwhile, for a given 3D fine-grained mask, 2D-based methods process each axial slice independently, whereas 3D-based methods holistically process and generate coherent 3D volumes. Consequently, 3D methods show better performance compared to 2D approaches. 

Furthermore, most existing methods primarily converge on large organs while exhibiting poor performance on fine structural details. Although MAISI produces satisfactory cardiac structures, it still exhibits intensity inaccuracies on small fine-grained details, resulting in inferior performance compared to our StructDiff in downstream pre-training. Through the mask-to-image correspondence brought by template-guided mechanism, our StructDiff enables the generation of images with enhanced topological consistency while achieving superior visualization of fine structural details. Furthermore, we invite two clinically experienced experts to evaluate and rank the synthesized data generated by various methods. Consequently, our proposed StructDiff achieves superior rankings and outstanding average scores, thereby highlighting the practical applicability of our methods in clinical scenarios. The details of the experts evaluation is provided in the supplementary material.

\vspace{-0.5mm}
\subsubsection{Quantitative Evaluation}
\vspace{-0.5mm}
As shown in \cref{tab1}, our proposed StructDiff achieves state-of-the-art performance on fine-grained cardiac image synthesis, which demonstrates the best structural fidelity and semantic representation comparable to real cardiac images. Generally speaking, most conditional methods show superior performance over unconditional methods. Although ControlNet and T2I-Adapter are 2D-based models, their integration with the generative capacity of diffusion models enables satisfactory performance on surrounding tissues. Moreover, SPADE exhibits minimal structural bias but significant intensity discrepancies, leading to relatively high SSIM and low RMSE. Meanwhile, MAISI employs 3D diffusion models for feature representation, achieving remarkable performance across four metrics. Notably, as shown in ``with surroundings'' and ``w/o surroundings'' parts in \cref{tab1}, a significant degradation is observed across all methods when surrounding tissues are removed from the reference. However, our StructDiff still produces outstanding results even in the absence of surrounding tissues. We attribute the robustness to the template-guided conditioning, which provides structural constraints for the generation process.

\begin{table}[t]
    \centering
    \caption{Influence on downstream segmentation pre-training when different number of templates are utilized during the generation.}
    \resizebox{\linewidth}{!}{
    \begin{tabular}{@{}c|c|cc|cc@{}}
    \toprule
        \multirow{2}*{\textbf{Group}} & \textbf{Number of} & \multicolumn{2}{c|}{\textbf{Cardiac Substructure}} & \multicolumn{2}{c}{\textbf{Coronary Artery}}\\
        \cline{3-6}
        ~ & \textbf{Template} & Dice $(\%)\uparrow$ & RDice $(\%)\uparrow$ & Dice $(\%)\uparrow$ & RDice $(\%)\uparrow$ \\
    \midrule
        \uppercase\expandafter{\romannumeral1} & 1 & 71.54$\pm$14.89 & 73.70$\pm$15.21 & 58.91$\pm$19.54 & 60.43$\pm$18.65\\
        \uppercase\expandafter{\romannumeral2} & 5 & 72.92$\pm$14.06 & 75.63$\pm$13.02& 60.23$\pm$17.25 & 62.72$\pm$16.81\\
        \uppercase\expandafter{\romannumeral3} & 10 & 75.32$\pm$13.11 & 78.22$\pm$11.48 & 62.14$\pm$16.42 & 66.45$\pm$14.93\\
        \uppercase\expandafter{\romannumeral4} & 20 &  \textbf{79.84$\pm$12.38} & \textbf{81.59$\pm$10.82} & \textbf{64.20$\pm$13.27} & \textbf{68.98$\pm$12.85}\\
        
    \bottomrule
    \end{tabular}}
    \label{tab4}
    \vspace{-4mm}
\end{table}

\vspace{-0.5mm}
\subsection{Downstream Segmentation Pre-training}
\vspace{-0.5mm}
\label{4.3}

To further evaluate the quality of synthesized data, we conduct experiments to assess their effectiveness in facilitating data-driven segmentation tasks, particularly when high-quality training data is scarce. As shown in \cref{tab2}, the ``Real-Only" baseline refers to the direct segmentation performance using limited real data samples without any pre-training, while the other methods incorporate 750 synthesized images generated by their respective models. Specifically, compared methods use only images as input and employ masks for supervision, whereas our StructDiff also introduces confidence maps, dynamically adjusting voxel-wise contributions of synthesized images to the supervision loss via the CAL strategy. For most methods, using synthetic images for pre-training brings Dice improvements, particularly when real data is limited. As listed in \cref{tab2}, pre-training with images generated by StructDiff combined with the CAL strategy achieves the best performance gains across all tasks for both nnU-Net and MedSAM.

Specifically, when using nnU-Net, StructDiff attains noticeable Dice improvements of 1.74$\%$, 3.88$\%$, 10.68$\%$, and 11.94$\%$ across the four datasets, respectively. Also, StructDiff helps MedSAM receive notable gains of 2.14$\%$, 5.71$\%$, 3.99$\%$, and 4.51$\%$. Notably,  MedSAM is a 2D framework and has higher tolerance for inter-slice discontinuity. herefore, images generated by 2D-based methods could achieve results comparable to ours on MedSAM.

\begin{figure}[t]
  \centering
   \includegraphics[width=\linewidth]{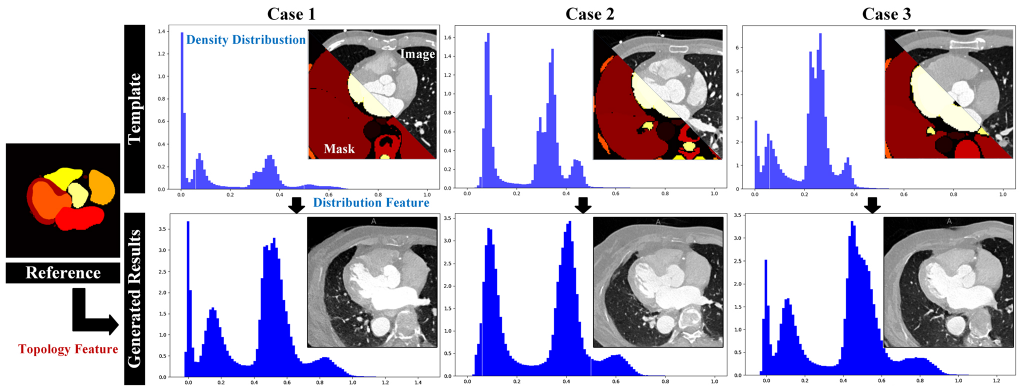}
   \caption{The generation results when different templates are applied on the same mask reference.}
   \label{fig5}
   \vspace{-4mm}
\end{figure}

\vspace{-1mm}
\subsection{Ablation Study and Analysis}
\vspace{-0.5mm}
\label{4.4}
To investigate the specific impacts of StructDiff on downstream pre-training and explain the roles of its main components, we conduct several experiments using nnU-Net\cite{isensee2021nnu} on our private cardiac dataset containing 288 cases. This dataset is carefully annotated by collaborating doctors, with more detailed and complete labels of the heart structure compared to publicly available datasets. We are committed to making these data open-source. The whole dataset is split into two parts, containing seven cardiac substructures as defined by MM-WHS\cite{zhuang2019mmwhs}, and the main branches of coronary arteries. More ablation studies on the publicly available dataset are provided in the supplementary material for reproducibility and verification.

\vspace{-0.5mm}
\subsubsection{The effectiveness of template}
As the core component within StructDiff, the template provides mask-to-image correspondence information, offering additional structural guidance and fine-grained topology. Here, we conduct experiments to verify its effectiveness. On the one hand, we keep the reference mask unchanged and use different image-mask templates to guide the generation process. As shown in \cref{fig5}, due to the difficulty of presenting 3D CTA images, we instead visualize the density distribution of both the template image and the synthesized result for comparison. It is obvious that different templates lead to significantly different results, increasing the complexity and richness of the synthesized dataset. Moreover, the density distribution of the generated image closely resembles its template, which clearly shows that the template plays a guiding role in the generation process.

To further investigate the effectiveness of the template, we conduct experiments where we pair 10 reference masks with 1, 5, 10, and 20 templates, respectively. Each group generates 100 synthetic images, which are then used for downstream segmentation training. Notably, all downstream models are trained under identical settings without fine-tuning on real data. As shown in \cref{tab4}, segmentation performance improves as the number of templates increases, demonstrating that templates have a positive impact on generation. More templates could provide more diverse yet reliable mask-to-image correspondence information, which leads to higher-quality generation results and better downstream performance.

\begin{table}[t]
    \centering
    \caption{Impact of StructDiff-synthesized images on downstream segmentation tasks under data scarcity.}
    \resizebox{\linewidth}{!}{
    \begin{tabular}{@{}c|cc|cc|cc@{}}
    \toprule
        \multirow{2}*{\textbf{Group}} & \multirow{2}*{\textbf{Real}} & \multirow{2}*{\textbf{Syn}} &  \multicolumn{2}{c|}{\textbf{Cardiac Substructure}} & \multicolumn{2}{c}{\textbf{Coronary Artery}}\\
        \cline{4-7}
        ~ & ~ & ~ & Dice $(\%)\uparrow$ & RDice $(\%)\uparrow$ & Dice $(\%)\uparrow$ & RDice $(\%)\uparrow$ \\
    \midrule
        \uppercase\expandafter{\romannumeral5} & 176 & 0 & 88.43$\pm$11.32 & 92.10$\pm$12.68 & 78.89$\pm$11.83 & 83.32$\pm$13.04\\
        \uppercase\expandafter{\romannumeral6} & 20 & 0 & 86.04$\pm$13.24 & 89.84$\pm$9.63& 72.37$\pm$13.49 & 79.54$\pm$12.81\\
        \uppercase\expandafter{\romannumeral7} & 20 & 400 & 88.62$\pm$12.57 & 91.75$\pm$9.79 & 78.91$\pm$11.39 & 83.23$\pm$12.76\\
        \uppercase\expandafter{\romannumeral8} & 20 & 1000 & \textbf{89.68$\pm$11.92} & \textbf{92.68$\pm$10.44} & \textbf{80.29$\pm$10.94} & \textbf{84.38$\pm$12.20}\\
    \bottomrule
    \end{tabular}}
    \label{tab3}
    \vspace{-4mm}
\end{table}

\vspace{-0.5mm}
\subsubsection{The effectiveness of main components}
\label{sec4.4.2}
On the one hand, we conduct experiments to verify the effectiveness of synthesized data in improving downstream pre-training. As listed in \cref{tab3}, groups \uppercase\expandafter{\romannumeral5} and \uppercase\expandafter{\romannumeral6} exhibit significant performance degradation when training data is reduced, particularly for coronary structures. 
Based on the group \uppercase\expandafter{\romannumeral6}, we gradually increase the number of synthesized samples. Specifically, when 400 synthetic images are introduced for pre-training followed by fine-tuning with limited real data, nnU-Net exhibits a noticeable recovery in accuracy (group \uppercase\expandafter{\romannumeral7}). As the number of synthesized samples increases to 1,000 (group \uppercase\expandafter{\romannumeral8}), nnU-Net continues to improve, ultimately surpassing the baseline that uses only real data. These results demonstrate that StructDiff can generate high-quality synthetic data that effectively enhances downstream performance. Moreover, generating 1,000 synthetic images corresponds to requiring 1,000 reference masks. However, coronary data are extremely challenging to annotate manually. Therefore, group \uppercase\expandafter{\romannumeral8} utilizes masks generated by our proposed MGM instead of manual annotations, further validating the effectiveness of MGM in enriching mask diversity and supporting scalable data generation.

On the other hand, we conduct an ablation study to evaluate the effectiveness of bidirectional training strategy, template condition, and CAL strategy. Notably, FID and LPIPS are utilized to assess the generation quality, while Avg Dice measures the impact of pre-training on downstream segmentation tasks. The results, listed in \cref{tab5}, show that all proposed components significantly enhance performance. Specifically, since CAL is only designed for pre-training stage, it does not influence the generation but substantially improves the accuracy of downstream segmentation task.

\begin{table}[t]
    \centering
    \caption{Ablation study of different components in StructDiff framework. Notably, ``One'' and ``Bi'' denote the presence or absence of a bidirectional training strategy in the StructDiff, while ``Norm'' refers to applying mask normalization alone and ``Temp'' indicates the proposed template-guided condition.}
    \resizebox{\linewidth}{!}{
    \begin{tabular}{@{}cc|cc|c|c|c|c@{}}
    \toprule
        \multicolumn{2}{c|}{\textbf{Direction}} & \multicolumn{2}{c|}{\textbf{Condition}}& \multirow{2}{*}{\textbf{CAL}} & \multirow{2}{*}{\textbf{FID$\downarrow$}} & \multirow{2}{*}{\textbf{LPIPS$\downarrow$}} & \multirow{2}{*}{\textbf{Avg Dice $(\%)\uparrow$}}\\
        \cline{1-4}
        \textbf{One} & \textbf{Bi} & \textbf{Norm} & \textbf{Temp}  & ~ & ~ & ~ & ~ \\
        \midrule
        $\surd$ & ~ & $\surd$ & ~ & ~ & 2.4239
 & 0.0842$\pm$0.0282 & 83.97$\pm$13.48 \\
        ~ & $\surd$ & $\surd$ & ~ & ~ & 2.3907 & 0.0811$\pm$0.0265  & 84.21$\pm$13.24 \\
        $\surd$ & ~ & ~ & $\surd$ & ~               & 1.6230 & 0.0673$\pm$0.0194 & 85.65$\pm$12.00 \\
        ~ & $\surd$ & ~ & $\surd$ & ~         & \textbf{1.5821}
 & \textbf{0.0541$\pm$0.0140} & 86.05$\pm$11.65 \\
        ~ & $\surd$ & ~ & $\surd$  & $\surd$   & \textbf{1.5821} & \textbf{0.0541$\pm$0.0140} & \textbf{87.27$\pm$10.31} \\
    \bottomrule
    \end{tabular}}
    \label{tab5}
    \vspace{-4mm}
\end{table}

\section{Conclusion}
\label{sec:conclusion}
In this work, we present a framework to address the scarcity of annotated medical images, especially for fine-grained anatomical regions with complex topology. To our best knowledge, it is the first work that enables the generation of fine-grained, high-quality medical images while preserving topological consistency. Our proposed StructDiff uses fine-grained masks as topological reference and introduces a paired image-mask template as additional guidance to provide explicit mask-to-image correspondence information. In addition, the training-free MGM enriches structural priors during generation and alleviates the scarcity of high-quality reference masks. Furthermore, to mitigate the influence of imperfect synthesized data in downstream pre-training tasks, CAL strategy incorporates SSV to estimate voxel-level uncertainty and adaptively adjust the contribution in the loss function. Experimental results demonstrate that StructDiff not only produces high-quality synthetic data but also significantly enhances downstream segmentation performance through CAL, highlighting its strong potential for advancing medical imaging analysis.
{
    \small
    \bibliographystyle{ieeenat_fullname}
    \bibliography{main}
}

\clearpage
\setcounter{page}{1}
\maketitlesupplementary

\section{More Details of Dataset and Metrics}
\label{sec:dm}
\subsection{Dataset}
\textbf{ImageCAS} \cite{zeng2023imagecas} is a large-scale coronary artery segmentation dataset containing 1,000 3D Computed Tomography Angiography (CTA) scans. It comprises data from 414 female and 586 male patients, with average ages of 59.98 and 57.68 years, respectively. The dataset provides detailed annotations for coronary arteries, including the Left Main (LM), Left Anterior Descending (LAD), Left Circumflex (LCX), Right Coronary Artery (RCA), First to Third Diagonal Branches (D1–D3), First to Third Obtuse Marginal Branches (OM1–OM3), Intermediate Branch (RI), Posterior Descending Artery (PDA), and Acute Marginal Branch (AM1), following the 17-segment nomenclature defined by the American Heart Association (AHA).

\noindent \textbf{ImageCHD} \cite{xu2020imagechd} contains 110 3D Computed Tomography (CT) scans covering a variety of congenital heart disease (CHD) types. Despite its moderate size, it is relatively substantial compared with existing medical imaging datasets. The dataset provides annotations for several cardiac structures, including the Left Ventricle (LV), Right Ventricle (RV), Left Atrium (LA), Right Atrium (RA), Myocardium (Myo), Aorta (Ao), and Pulmonary Artery (PA).

\noindent \textbf{ImageALCAPA} \cite{zeng2022imagealcapa} comprises 30 3D CTA scans acquired using a SOMATOM Definition Flash CT system. All scans are preoperative CTA images of patients with ALCAPA (Anomalous Left Coronary Artery from the Pulmonary Artery). The dataset includes annotations for multiple anatomical structures, such as the Myocardium (Myo), Left Ventricle (LV), Right Ventricle (RV), Pulmonary Artery (PA), Aorta (Ao), Left Coronary Artery (LCA), and Right Coronary Artery (RCA).

\noindent \textbf{ASOCA} \cite{gharleghi2022automated} provides a training set of 40 cardiac CTA (CCTA) scans depicting coronary arteries. The dataset includes 20 healthy subjects and 20 patients diagnosed with coronary artery disease. The annotations cover the left and right coronary arteries and their major branches, including the Left Anterior Descending (LAD), Left Circumflex (LCX), Septal, Diagonal, Obtuse Marginal, and Ramus Intermedius segments (when present).

\noindent \textbf{MM-WHS} \cite{zhuang2019mmwhs} comprises 120 multimodal cardiac images, including 60 cardiac CT/CTA scans and 60 cardiac MRI scans. These images cover the whole heart and its major substructures and were collected in real clinical settings for diagnostic purposes. The training set provides manual annotations for seven key cardiac structures: the Left Ventricle (LV), Right Ventricle (RV), Left Atrium (LA), Right Atrium (RA), Myocardium (Myo), Ascending Aorta (Ao), and Pulmonary Artery (PA).

\begin{figure*}[t]
  \centering
  \resizebox{\textwidth}{!}{
   \includegraphics[width=\textwidth]{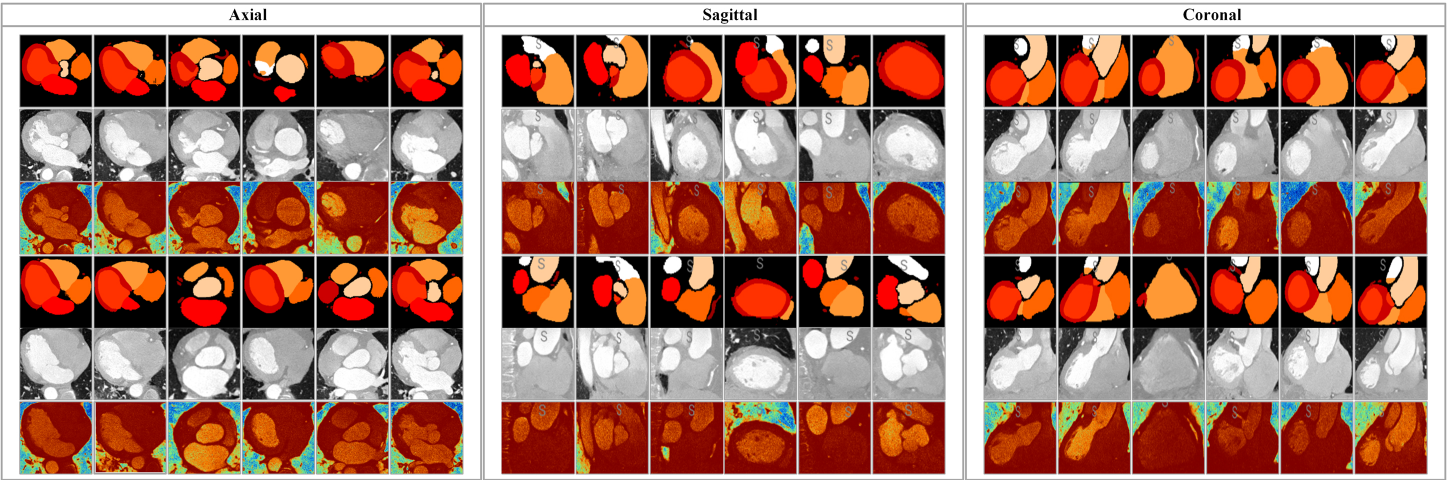}}
   \caption{Additional visualization results of StructDiff. Three distinct perspectives are presented to demonstrate the given mask and corresponding generated images as well as confidence maps.}
   \label{A1}
\end{figure*}

\subsection{Metrics}
We evaluate image quality using four metrics \cite{guo2024maisi}: Structural Similarity Index (SSIM), Root Mean Square Error (RMSE), Fréchet Inception Distance (FID), and Learned Perceptual Image Patch Similarity (LPIPS). Specifically, SSIM, RMSE, and LPIPS are computed for each real–synthetic image pair under the same mask, whereas FID is computed between the full sets of real and synthesized images to assess the global distribution discrepancy. Since unconditional methods do not rely on masks, only FID is applicable to them. During pre-training, we additionally adopt Dice and RDice \cite{qi2021examinee} as evaluation metrics.

\noindent \textbf{SSIM} 
(Structural Similarity Index) is an objective metric that measures image quality from three perceptual dimensions: luminance, contrast, and structural consistency. For a given mask $\boldsymbol{M}$, SSIM is computed using \cref{eq10} between the generated image $\boldsymbol{X}$ and the corresponding ground-truth image $\boldsymbol{Y}$. The metric ranges from 0 to 1, with higher values indicating greater similarity.
\begin{equation}
\resizebox{0.85\linewidth}{!}{$
\operatorname{SSIM}(\boldsymbol{X}, \boldsymbol{Y})=\frac{\left(2 \mu_{\boldsymbol{X}}\mu_{\boldsymbol{Y}}+{C}_{1}\right)\left(2 \sigma_{\boldsymbol{X}\boldsymbol{Y}}^2+{C}_{2}\right)}{\left(\mu_{\boldsymbol{X}}^{2}+\mu_{\boldsymbol{Y}}^{2}+{C}_{1}\right)\left(\sigma_{\boldsymbol{X}}^{2}+\sigma_{\boldsymbol{Y}}^{2}+{C}_{2}\right)}$},
\label{eq10}
\end{equation}
where $\mu_{\boldsymbol{X}}$ and $\mu_{\boldsymbol{Y}}$ denote the mean intensities of $\boldsymbol{X}$ and $\boldsymbol{Y}$, respectively. $\sigma_{\boldsymbol{X}\boldsymbol{Y}}^2$ denotes their covariance, and $\sigma_{\boldsymbol{X}}^2$ and $\sigma_{\boldsymbol{Y}}^2$ denote the variances of $\boldsymbol{X}$ and $\boldsymbol{Y}$, respectively.

\noindent \textbf{RMSE}
(Root Mean Square Error) is a commonly used statistical metric that quantifies differences between predicted and ground-truth values. A smaller RMSE means a closer match between the two. The RMSE is computed as: 
\begin{equation}
\resizebox{0.85\linewidth}{!}{$
\mathrm{RMSE}(\boldsymbol{X},\boldsymbol{Y})=\sqrt{\frac{1}{h\cdot w\cdot d} \sum_{i=1}^h\sum_{j=1}^w\sum_{k=1}^d\left(\boldsymbol{X}_{i,j,k}-\boldsymbol{Y}_{i,j,k}\right)^2},$}
\label{eq11}
\end{equation}
\noindent where $h$, $w$, and $d$ denote the height, width, and depth of the image, respectively.

\noindent \textbf{FID}
(Fréchet Inception Distance) measures the quality of generated images by comparing the feature distributions of real and synthesized image sets. A lower FID indicates a smaller distributional discrepancy between the two sets in the feature space. It is defined as:
\begin{equation}
  \begin{split}
\mathrm{FID}(\boldsymbol{\mathrm{X}},\boldsymbol{\mathrm{Y}}) &=\left\|\boldsymbol\mu_{r}-\boldsymbol\mu_{g}\right\|^2\\
&+\operatorname{Tr}\left(\boldsymbol\Sigma_{r}+\boldsymbol\Sigma_{g}-2\left(\boldsymbol\Sigma_{r} \boldsymbol\Sigma_{g}\right)^\frac{1}{2}\right),
  \end{split}
\label{eq12}
\end{equation}
where $\operatorname{Tr}(\cdot)$ denotes the trace of a matrix. $\boldsymbol{\mu}$ and $\boldsymbol{\Sigma}$ represent the mean and covariance of the extracted features, respectively. $\boldsymbol{X}$ and $\boldsymbol{Y}$ denote the sets of generated and ground-truth images, while the subscripts $r$ and $g$ refer to the corresponding real and generated feature distributions.

\noindent \textbf{LPIPS}
(Learned Perceptual Image Patch Similarity) is a perceptual metric designed to assess the similarity between images by approximating human visual judgments. It measures differences in deep feature space rather than relying solely on pixel-level comparisons. LPIPS is computed as:
\begin{equation}
\operatorname{LPIPS}(\boldsymbol{X}, \boldsymbol{Y})=\sum_{i=1}^L w_i \cdot \frac{1}{N_i}\left\|f_i(\boldsymbol{X})-f_i(\boldsymbol{Y})\right\|_2^2,
\label{eq13}
\end{equation}
where $f_i$ denotes the feature map extracted from the $i^{\text{th}}$ layer of the network, and $N_i$ is the number of elements in that feature map. The term $w_i$ represents the learned weight associated with the $i^{\text{th}}$ layer.

\noindent \textbf{Dice}
(Dice Coefficient) is a standard metric used to evaluate the similarity between two sets and is widely applied in image segmentation tasks. Given an input image $\boldsymbol{I}$ and its corresponding ground-truth mask $\boldsymbol{Q}$, the segmentation model produces a predicted mask $\boldsymbol{P}$. The Dice coefficient is computed using two sets of pixels as:
\begin{equation}
\operatorname{Dice}(\boldsymbol{P}, \boldsymbol{Q})=\frac{2\vert\boldsymbol{P}\cap\boldsymbol{Q}\vert}{\vert\boldsymbol{P}\vert+\vert\boldsymbol{Q}\vert}.
\label{eq14}
\end{equation}

\noindent \textbf{RDice}
(Relative Dice Coefficient) is a variant of the Dice coefficient designed to enhance robustness when segmenting small or fine anatomical structures. The ground-truth mask $\boldsymbol{Q}$ is first processed using morphological dilation to obtain an expanded mask $\boldsymbol{Q}'$. The intersection of $\boldsymbol{Q}'$ with the predicted mask $\boldsymbol{P}$ yields $\boldsymbol{P}'$. RDice is then defined as:
\begin{equation}
\operatorname{RDice}(\boldsymbol{P}, \boldsymbol{Q})=\frac{2\vert\boldsymbol{P}'\cap\boldsymbol{Q}\vert}{\vert\boldsymbol{P}'\vert+\vert\boldsymbol{Q}\vert},
\label{eq15}
\end{equation}

\section{More Comparison Results}
\subsection{Synthetic Images of StructDiff}
In \cref{A1}, we present additional synthesized cardiac images generated by our StructDiff. The results show that StructDiff preserves complete structural continuity across axial, sagittal, and coronal planes, without introducing anatomical discontinuities. At the same time, the synthesized images exhibit intensity distributions that closely match those of real images. The confidence maps in \cref{A1} further reveal distinct reliability patterns: fine-grained masked regions display higher confidence values, while unmasked areas exhibit reduced certainty. When used for downstream segmentation pre-training, these confidence maps serve as dynamic weights that adaptively modulate voxel-wise contributions, helping mitigate the influence of imperfections in synthesized data and improving its overall usability.

\subsection{Segmentation Evaluation and Feature Distribution}
\begin{figure*}[t]
  \centering
  \resizebox{\textwidth}{!}{
   \includegraphics[width=\textwidth]{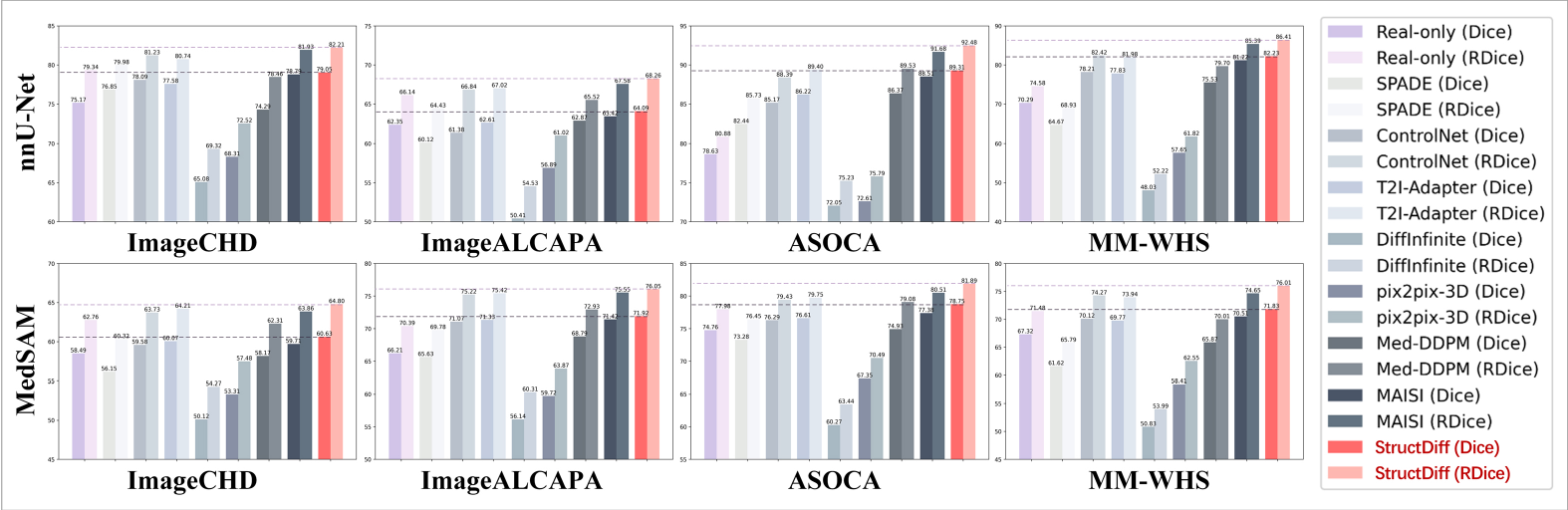}}
   \caption{Comparative bar plots of pre-training performance metrics across four public datasets for different methods. Downstream segmentation models pre-trains using StructDiff-synthesized data and subsequently fine-tunes with limited real data, achieving significant improvements in both Dice and RDice metrics.}
   \label{A2}
\end{figure*}

\begin{figure*}[t]
  \centering
  \resizebox{\textwidth}{!}{
   \includegraphics[width=\textwidth]{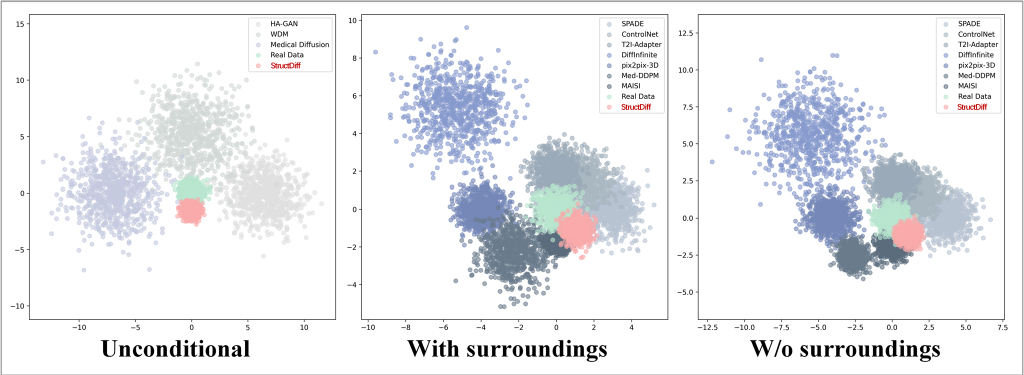}}
   \caption{Scatter plots comparing feature distributions of data synthesized by unconditional versus conditional methods.}
   \label{A5}
\end{figure*}

\Cref{A2} provides a visual representation of \Cref{tab2} from the main paper using bar chart comparisons. These results clearly illustrate that our synthesized images effectively support downstream pre-training when real data is limited. Compared with other methods, StructDiff achieves larger performance gains in pre-training tasks, particularly under data-scarce settings.

To further assess the realism of synthesized images, we apply t-SNE \cite{he2023geometric} to visualize the feature distributions of 750 synthesized samples generated by different methods in feature space. The feature representations are extracted from the intermediate hidden layer of an nnU-Net model pre-trained on the ImageCAS \cite{zeng2023imagecas} training set. The feature distribution of real images serves as the baseline for comparison. As shown in \cref{A5}, we group the visualized distributions into three categories—unconditional, with surroundings, and without surroundings—consistent with the grouping in \Cref{tab1} of the main paper.

\begin{figure*}[!t]
  \centering
  \resizebox{\textwidth}{!}{
   \includegraphics[width=\textwidth]{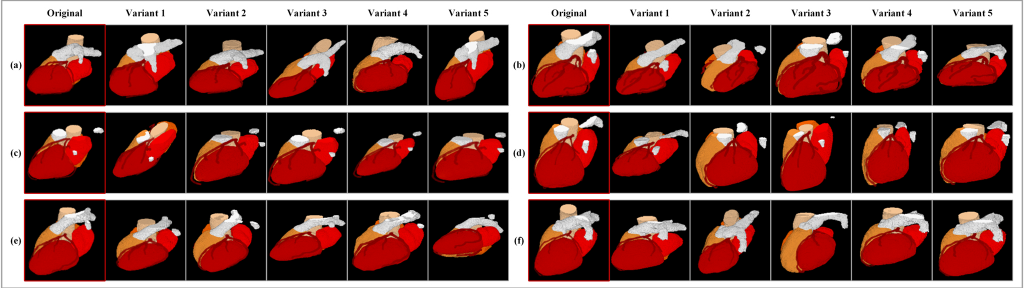}}
   \caption{The topologically consistent and morphologically heterogeneous masks synthesized by MGM.}
   \label{A6}
\end{figure*}

Compared with conditional methods, unconditional approaches \cite{sun2022hierarchical, friedrich2024wdm, khader2023denoising} exhibit feature embeddings that deviate substantially from those of real data (green) and show large inter-class variance, resulting in highly scattered clusters. This phenomenon stems from the absence of topological constraints normally provided by structural masks. Among the conditional methods, DiffInfinite \cite{aversa2023diffinfinite} shows poor spatial adherence to mask guidance and large intensity inconsistencies. In contrast, other conditional methods yield feature embeddings that lie closer to the real data distribution (green), with fewer outliers due to the topological constraints imposed by the conditional mask. Notably, the features of samples generated by our StructDiff (red) form the most compact clusters and lie closest to real images, further supporting our qualitative observations. MAISI \cite{guo2024maisi} achieves comparable performance in the ``with surroundings" setting but exhibits noticeable distribution drift in the ``without surroundings" setting, suggesting that the absence of surrounding tissue context leads to increased dispersion and greater deviation in feature space for most methods.

Overall, these visualization results confirm that StructDiff synthesizes fine-grained 3D cardiac images whose feature distributions closely align with those of real data. This not only verifies the geometric fidelity of our synthesized images but also ensures semantically coherent representations, thereby significantly enhancing the effectiveness of downstream pre-training tasks.

\begin{table}[t]
    \centering
    \footnotesize
    \caption{Ablation study for validating whether the deformation of the mask by MGM impairs the guiding capability of the mask.}
    \resizebox{\linewidth}{!}{
    \begin{tabular}{c|c|c|cc|cc}
    \toprule
         \multirow{2}{*}{\textbf{Group}} & \textbf{Original} & \textbf{Deformed Mask} &  \multicolumn{2}{c|}{\textbf{ImageCAS}}&  \multicolumn{2}{c}{\textbf{MM-WHS}}\\
         \cline{4-7}
         & \textbf{Mask} & \textbf{using MGM} & Dice($\%$)$\uparrow$ & RDice($\%$)$\uparrow$ & Dice($\%$)$\uparrow$ & RDice($\%$)$\uparrow$\\
    \midrule
        \uppercase\expandafter{\romannumeral9} & 50 & 0  & 76.25$\pm$8.74 & 79.57$\pm$7.06 & 89.09$\pm$2.53 & 92.35$\pm$2.67 \\
        \uppercase\expandafter{\romannumeral10} & 0 & 50 & \textbf{76.73$\pm$8.61} & \textbf{79.82$\pm$6.98} & 88.23$\pm$2.80 & 92.04$\pm$2.78 \\
        \uppercase\expandafter{\romannumeral11} & 25 & 25 & 76.37$\pm$8.66 & 79.78$\pm$7.11 & \textbf{89.31$\pm$2.76} & \textbf{92.48$\pm$2.83}\\
    \bottomrule
    \end{tabular}}
    \label{tab:A1}
\end{table}

\subsection{The MGM to Generate Different Masks}
To examine the structural validity of reference masks generated by MGM, we produce multiple variants for each original mask and evaluate their morphological diversity and topological consistency. As shown in \cref{A6}, six representative cases are selected, and five variants are generated for each original mask using a combination of affine transformations (rotation, scaling, shearing, and translation) and non-rigid deformations. These results indicate that the derived masks preserve essential topological relationships (e.g., positional alignment between coronary arteries and ventricles). In contrast to applying deformations directly on images, which often introduce localized inappropriate expansions or compressions that violate global anatomical coherence, our MGM applies deformations exclusively to fine-grained masks. This design ensures that images synthesized using these masks as structural inputs maintain anatomically consistent surrounding tissues. At the same time, randomized deformation parameters introduce sufficient morphological variability while preserving topology. Consequently, MGM-generated masks achieve both morphological diversity and topological fidelity, effectively supporting precise structure-aware synthesis within the StructDiff framework.

Moreover, as presented in \Cref{tab:A1}, we conduct experiments on two public datasets, ImageCAS and MM-WHS, to further demonstrate the effectiveness of MGM. All experiments in \Cref{tab:A1} utilize 50 original masks and 50 reference masks to produce 750 synthetic images, which are subsequently used to pre-train the nnU-Net. After pre-training, nnU-Net is fine-tuned using 20 ImageCAS cases for coronary artery segmentation and 20 MM-WHS cases for cardiac substructure segmentation. The key difference between Groups \uppercase\expandafter{\romannumeral9} and \uppercase\expandafter{\romannumeral10} lies in the source of the reference masks: Group \uppercase\expandafter{\romannumeral9} uses the original, non-deformed reference masks, whereas all reference masks in Group \uppercase\expandafter{\romannumeral10} are generated via MGM. Group \uppercase\expandafter{\romannumeral11}, on the other hand, uses a mixture of both original and MGM-deformed masks randomly selected from the previous two groups. These experiments are designed to assess the robustness and reliability of MGM-generated reference masks. The results show that both original and MGM-deformed masks yield comparable performance, indicating that MGM-generated masks can serve as effective alternatives to non-deformed reference masks and thereby alleviate the scarcity of high-quality reference annotations. Furthermore, the experiments in \Cref{tab:A1} confirm that the MGM deformation process preserves the topological structures essential for effective structural conditioning.

\subsection{Manual Evaluation from Clinical Experts}
\begin{figure*}[t]
  \centering
  \resizebox{\textwidth}{!}{
   \includegraphics[width=\textwidth]{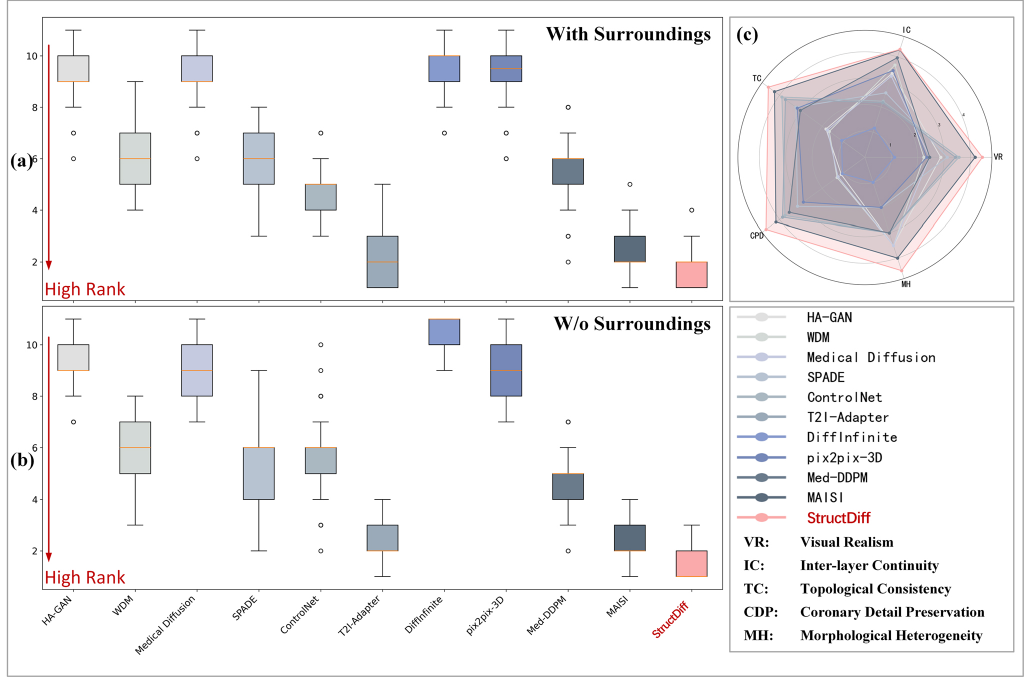}}
   \caption{Visualization of clinical experts scoring and ranking results. (a) and (b) present box plots of ranking statistics (mean ± standard deviation), demonstrating consistent top-tier rankings for StructDiff. (c) employs a radar chart to compare synthesized images across methods, showing that StructDiff-generated images achieve significantly higher expert scores compared to other methods.}
   \label{A7}
   \vspace{-3mm}
\end{figure*}

In addition to quantitative evaluations of synthesized image quality, we further conduct a comprehensive expert assessment by inviting two clinicians with extensive cardiovascular imaging experience to systematically score and rank the synthetic images generated by different models. As illustrated in \cref{A7}, synthesized images from 11 conditional and unconditional methods are selected for manual evaluation. For each method, 200 images synthesized with surrounding structures and 200 images synthesized without surrounding structures are included. Images generated from the same reference across all 11 models are grouped into an “evaluation bag.” Thus, each bag contains 11 synthesized images, one from each model, and the clinicians are required to independently score every image. The evaluation follows five key criteria: Visual Realism, Inter-layer Continuity, Topological Consistency, Coronary Detail Preservation, and Morphological Heterogeneity. Within each evaluation bag, the 11 models are ranked based on their scores, where a rank of 1 indicates the best performance and 11 indicates the worst. Accordingly, lower rank values correspond to better overall model performance.

For each model, we compute the mean and variance of its ranks across all evaluation bags and visualize the results using box plots. As shown in \cref{A7}(a) and \cref{A7}(b), a lower box position and a shorter box height indicate that the corresponding model achieves consistently better rankings across different references. Overall, the proposed StructDiff exhibits consistently superior performance. Furthermore, \cref{A7}(c) presents a radar chart comparing all models across the five evaluation criteria. StructDiff forms the largest and most balanced pentagon, highlighting its leading performance across all qualitative dimensions.

\begin{table}[t]
    \centering
    \caption{Runtime comparison with diffusion-based models.}
    \begin{tabular}{l|ccc}
        \hline
         \footnotesize\textbf{Model}& \footnotesize\textbf{SSIM}$\uparrow$ & \footnotesize\textbf{FID}$\downarrow$ &\footnotesize\textbf{Time(s)}$\downarrow$ \\
         \hline
         \footnotesize ControlNet&  \footnotesize 0.8739$\pm$0.0197& \footnotesize 2.0073 & \footnotesize 618$\pm$21\\
         \footnotesize MAISI&  \footnotesize 0.9110$\pm$0.0158& \footnotesize 1.6881& \footnotesize 873$\pm$46\\
         \footnotesize\textbf{StructDiff (ours)}  & \footnotesize\textbf{0.9243$\pm$0.0165} & \footnotesize\textbf{1.5821} & \footnotesize \textbf{312$\pm$27}\\
         \hline
    \end{tabular}
    \label{tab1}
    \vspace{-5mm}
\end{table}

\subsection{Runtime Analyses}
To further demonstrate the effectiveness of StructDiff, we compare it with two diffusion-based models—ControlNet and MAISI—which represent state-of-the-art approaches and have attracted significant attention in this field. As reported in \Cref{tab1}, StructDiff achieves notably faster inference speed than both competitors. It is worth noting that the synthesized images in our task are intended for offline use rather than real-time intraoperative applications. Therefore, although StructDiff exhibits clear advantages in efficiency, inference speed is not the primary performance determinant for this task. Nonetheless, the improved efficiency further highlights the practicality and scalability of our framework.

\begin{table}
    \centering
    \footnotesize
    \caption{Impact of StructDiff-synthesized images on downstream segmentation tasks under data scarcity on public datasets.}
    \resizebox{\linewidth}{!}{
    \begin{tabular}{c|c|c|cc|cc}
    \toprule
        \multirow{2}{*}{\textbf{Group}} & \multirow{2}{*}{\textbf{Real}} & \multirow{2}{*}{\textbf{Syn}} &  \multicolumn{2}{c|}{\textbf{ImageCAS}}&  \multicolumn{2}{c}{\textbf{MM-WHS}}\\
    \cline{4-7}
        &  &  & Dice($\%$)$\uparrow$ & RDice($\%$)$\uparrow$ & Dice($\%$)$\uparrow$ & RDice($\%$)$\uparrow$\\
    \midrule
        \uppercase\expandafter{\romannumeral12} & 20 & 0 & 74.73$\pm$8.84 & 78.59$\pm$7.61 & 85.49$\pm$4.83 & 87.33$\pm$3.56\\
        \uppercase\expandafter{\romannumeral13} & 20 & 400 & 79.04$\pm$8.18 & 82.38$\pm$6.87 & 87.24$\pm$4.66 & 89.19$\pm$3.05\\
        \uppercase\expandafter{\romannumeral14} & 20 & 1000 & \textbf{81.44$\pm$7.58} & \textbf{84.09$\pm$6.26} & \textbf{89.98$\pm$3.92} & \textbf{91.84$\pm$2.92}\\
    \bottomrule
    \end{tabular}}
    \label{tab:A2}
    \vspace{-5mm}
\end{table}

\begin{table}[t]
    \centering
    \caption{Ablation study of different components in StructDiff framework on public datasets.}
    \resizebox{\linewidth}{!}{
    \begin{tabular}{@{}cc|cc|c|c|c|cc@{}}
    \toprule
        \multicolumn{2}{c|}{\textbf{Direction}} & \multicolumn{2}{c|}{\textbf{Condition}}& \multirow{2}{*}{\textbf{CAL}} & \multirow{2}{*}{\textbf{FID$\downarrow$}} & \multirow{2}{*}{\textbf{LPIPS$\downarrow$}} & \textbf{ImageCAS} & \textbf{MM-WHS}\\
        \cline{1-4}
        \textbf{One} & \textbf{Bi} & \textbf{Norm} & \textbf{Temp}  & ~ & ~ & ~ & Dice $(\%)\uparrow$ & Dice $(\%)\uparrow$\\
        \midrule
        $\surd$ & ~ & $\surd$ & ~ & ~ & 2.4239 & 0.0842$\pm$0.0282 & 73.57$\pm$9.16 & 85.81$\pm$5.03\\
        ~ & $\surd$ & $\surd$ & ~ & ~ & 2.3907 & 0.0811$\pm$0.0265  & 74.29$\pm$8.97 & 86.43$\pm$4.82\\
        $\surd$ & ~ & ~ & $\surd$ & ~ & 1.6230 & 0.0673$\pm$0.0194 & 76.84$\pm$8.62 & 87.95$\pm$4.34\\
        ~ & $\surd$ & ~ & $\surd$ & ~ & \textbf{1.5821} & \textbf{0.0541$\pm$0.0140} & 77.33$\pm$8.56 & 88.48$\pm$4.11\\
        ~ & $\surd$ & ~ & $\surd$  & $\surd$   & \textbf{1.5821} & \textbf{0.0541$\pm$0.0140} & \textbf{80.75$\pm$8.07} & \textbf{89.13$\pm$4.08}\\
    \bottomrule
    \end{tabular}}
    \label{tab:A3}
    \vspace{-4mm}
\end{table}

\subsection{Extensions of Ablation Studies}
To further support reproducibility and independent verification, we provide additional ablation studies on publicly available datasets in \Cref{tab:A2} and \Cref{tab:A3}.

As shown in \Cref{tab:A2}, Groups \uppercase\expandafter{\romannumeral12}–\uppercase\expandafter{\romannumeral14} serve as an extension and supplement to \Cref{tab4} in the main paper. The results consistently demonstrate that the synthetic data generated by StructDiff, together with the CAL strategy, provides an effective prior for downstream segmentation under data-scarce conditions. Using the same pre-training settings as in \Cref{tab4}, we fine-tune nnU-Net on 20 cases from ImageCAS and MM-WHS, respectively, and obtain conclusions aligned with those in \Cref{sec4.4.2} of the main paper. It validates the effectiveness of our design on both private and public datasets, confirming its practical applicability.

As shown in \Cref{tab:A3}, we further evaluate the effectiveness of the CAL strategy on public datasets. Since the data used during the generation stage remains unchanged, we directly retain the generation metrics reported in \Cref{tab5} of the main paper. Following the same experimental setup as \Cref{tab5}, we pre-train and fine-tune nnU-Net on the public datasets. For ImageCAS, we randomly select 50 samples for fine-tuning and 200 samples for testing, maintaining a 1:4 data split ratio. The experimental results lead to the same conclusions as in \Cref{sec4.4.2}, further validating both the module designs of StructDiff and the effectiveness of the CAL strategy on public datasets.

\end{document}